\documentclass{article}

\usepackage{arxiv}

\usepackage[utf8]{inputenc} 
\usepackage[T1]{fontenc}    
\usepackage{hyperref}       
\usepackage{url}            
\usepackage{booktabs}       
\usepackage{amsfonts}       
\usepackage{nicefrac}       
\usepackage{microtype}      
\usepackage{lipsum}
\usepackage{graphicx}
\usepackage{fancyhdr}
\usepackage{enumitem}
\usepackage{multirow}
\usepackage{algpseudocode}
\usepackage[most]{tcolorbox}
\usepackage{booktabs}
\usepackage{enumitem}
\usepackage{placeins}
\usepackage{xurl}
\graphicspath{ {./images/} }

\usepackage{tabularx}
\usepackage{array}
\usepackage{makecell}

\newcolumntype{Y}{>{\centering\arraybackslash}X}

\title{Can ChatGPT Generate Realistic Synthetic System Requirement Specifications? Results of a Case Study}

\author{
  Alex R. Mattukat \\
  Research Group Software Construction\\
  RWTH Aachen University\\
  Ahornstraße 55 \\
  Aachen, Germany \\
  \texttt{mattukat@swc.rwth-aachen.de} 
  \And
  Florian M. Braun \\
  Research Group Software Construction\\
  RWTH Aachen University\\
  Ahornstraße 55 \\
  Aachen, Germany \\
  \texttt{florian.maximilian.braun@rwth-aachen.de}
  \And
  Horst Lichter \\
  Research Group Software Construction\\
  RWTH Aachen University\\
  Ahornstraße 55 \\
  Aachen, Germany \\
  \texttt{lichter@swc.rwth-aachen.de}
  \\
}

\begin{document}
\tiny
This is the \textbf{peer-reviewed, accepted version} of a paper, including minor changes to content that was anonymized for double-blind review. It will appear in the proceedings of the 
21st International Conference on Evaluation of Novel Approaches of Software Engineering (\textbf{ENASE 2026}). The final published version will be available from Science and Technology Publications (SCITEPRESS).
© 2026 SCITEPRESS. Personal use of this material is permitted. Permission from SCITEPRESS must be obtained for all other uses, in any current or future media, including reprinting or republishing this material for advertising or promotional purposes, creating new collective works, resale or redistribution to servers or lists, or reuse of any copyrighted component of this work in other works.
\normalsize
\maketitle
\thispagestyle{fancy}
\begin{abstract}
System requirement specifications (SyRSs) are central, natural-language (NL) artifacts. Access to real SyRS for research purposes is highly valuable but limited by proprietary restrictions or confidentiality concerns. Generating synthetic SyRSs (SSyRSs) can address this scarcity. Black-box large language models (LLMs) such as ChatGPT offer compelling generation capabilities by providing easy access to NL generation functions without requiring access to real data. However, LLMs suffer from hallucinations and overconfidence, which pose major challenges in their use. We designed an exploratory study to investigate whether, despite these challenges, we can generate realistic SSyRSs with ChatGPT without having access to real SyRSs. Using a systematic approach that leverages prompt patterns, LLM-based quality assessments, and iterative prompt refinements, we generated 300 SSyRSs across 10 industries with ChatGPT. The results were evaluated using cross-model checks and an expert study, with n=87 submitted surveys.  62\% of experts considered the SSyRSs to be realistic. However, in-depth examination revealed contradictory statements and deficiencies. Overall, we were able to generate realistic SSyRSs to a certain extent with ChatGPT, but LLM-based quality assessments cannot fully replace thorough expert evaluations. This paper presents the methodology and results of our study and discusses the key insights we obtained.
\end{abstract}


\newtcolorbox{mainBox}[1]{%
		enhanced,
		colback=white!90!black,
		colframe=darkgray,
		colbacktitle=darkgray,
		coltitle=white,
		fonttitle=\bfseries,
		title=#1,
		rounded corners,
		boxrule=3pt,
		drop shadow=darkgray,
		width=\textwidth,
	}

\section{\uppercase{Introduction}}
\label{sec:introduction}

The System Requirements Specification (SyRS) is a central artifact of Requirements Engineering (RE). According to ISO/IEC/IEEE 29148:2018 \cite{ISO.29148}, it is a structured collection of the functional and non-functional requirements, design constraints, and other related information for the system, its operational environments, and external interfaces. They are mostly documented in natural language (NL), as it is more accessible to all stakeholders than formal languages \cite{Kassab.2014}. Access to real SyRSs is of high value for many research activities in software engineering (SE) research, such as developing methods and tools for elicitation, and validation of requirements \cite{Mehraj.2024}, automated testing and test case generation \cite{Gurbuz.2018}, tool benchmarking and evaluation \cite{Ferrari.2017_2}, and more. However, access to real SyRS is limited due to proprietary restrictions, confidentiality concerns, or simply unavailability \cite{Ferrari.2017_2}. One way to alleviate this is to generate realistic synthetic SyRSs (SSyRSs) that model real SyRSs realistically. 

Large Language Models (LLMs) offer compelling opportunities for generating SSyRSs, due to their capabilities in natural language processing (NLP) \cite{Bhattacharya.2024}. Black-box LLMs such as ChatGPT are particularly attractive. They offer fast and easy access to generation capabilities for NL artifacts, as they neither require initial training by the user nor access to real data, and are available via directly accessible APIs. However, Mehraj et al. \cite{Mehraj.2024} point out that research in the area of LLMs in RE primarily focuses on identifying and analyzing requirements. Hence, the feasibility of LLMs for generating synthetic requirement artifacts remains underexplored. 

However, while LLMs seem attractive for this use case, they come with major challenges that severely limit their usability for generating NL artifacts. \textit{Hallucination} is the phenomenon that LLMs unknowingly generate data that is nonsensical, unfaithful to the input source, or unverifiable. \textit{Overconfidence} describes the tendency of LLMs to present information with high certainty, regardless of whether the content is factually correct or verifiable. Both phenomena interplay dangerously: First, an LLM generates incorrect data by hallucinating. Then, it presents the incorrect data in NL in an overly confident manner, which easily convinces the user that the presented data is correct \cite{Huang.2025}.

Moreover, generating NL artifacts in expert domains, which is the case for SSyRSs, reveals a particular challenge: black-box LLMs are mainly trained on internet data \cite{Nie.2024}, but evaluated expert domain knowledge is shared far less over the internet than common knowledge. This results in significantly smaller---and possibly incorrect---training datasets. As LLMs tend to generate wrong data than admit to not knowing something \cite{Cheng.2024}, this increases the risk of hallucinations. While there exist mitigation strategies for hallucination, they rely on access to internal model parameters, which black-box LLMs do not provide \cite{Huang.2025}. As a result, alternative methods, such as output verification with external knowledge, auxiliary fact-checking models, or structured prompt engineering, must be employed when using black-box LLM \cite{Huang.2025,Manakul.2023}.

\textit{Contributions}: This paper presents an exploratory study that investigates whether black-box LLMs can be used to generate realistic SSyRSs. Being exploratory, the goal of our study was not to optimize the degree of realism of the generated data by any means necessary. Instead, we wanted to investigate whether we can leverage black-box LLMs---due to their easy access and strong NL capabilities---for generating NL expert data in contexts where there is no access to real, high-quality evaluated data and domain experts, as is often the case in the context of SyRS, especially in research contexts. We selected ChatGPT because it is a widely used black-box LLM and offers a good balance of performance, accessibility, and costs. The following research question concludes our main objective:

\vspace{10pt}

\noindent \textbf{RQ1:} \textit{Can ChatGPT be used to generate and assure realistic SSyRSs without access to real SyRSs and domain expert involvement?}

\vspace{10pt}


To give an answer, we must define suitable quality properties and ensure that these adequately describe the quality of the generated artifacts in terms of their intended use, i.e., that the SSyRSs can be used as substitutes for real SyRSs. This is described by the second research question:

\vspace{10pt}

\noindent \textbf{RQ2:} \textit{What quality properties can be used to describe the suitability of SSyRSs to be used in RE research as substitutes for real SyRSs?}

\vspace{10pt}


If we can generate sufficiently suitable SSyRSs with ChatGPT in terms of these quality characteristics, it remains to answer whether human experts consider the SSyRSs to be realistic enough to replace real SyRSs:

\vspace{10pt}

\noindent \textbf{RQ3:} \textit{Are assessments of SSyRSs made by human experts in line with ChatGPT-based quality assessments?}

\vspace{10pt}

The paper is structured as follows: Section 2 presents our methodology and our generation and assessment approach. Section 3 presents the statistics and an example SSyRSs of our final dataset. Section 4 presents the expert evaluation study we conducted. The discussion follows in section 5. Section 6 briefly discusses related work. Section 7 concludes the paper.

\section{\uppercase{Research Design}}
Design science research (DSR) is a widely accepted approach in information systems and SE research \cite{Peffers.2007}. As the aim of our paper is to develop a generational approach for SSyRSs, our study follows DSR. We already presented the identified problem and motivation in \autoref{sec:introduction}. Based on a targeted literature review to gain an overview of the field, we have discovered compelling use cases for the application of black-box LLMs for the generation, evaluation, and refinement of SSyRSs \cite{Gu.2025,Plaat.2024}. Due to the limited access to real SyRSs, these insights helped us formulate the research objective for this study.:

\vspace{5pt}

\noindent \textit{Examine whether ChatGPT can generate realistic SSyRSs for various industry domains (\textit{domains}) without providing real SyRSs or involving domain experts.}

\vspace{5pt}

\noindent To use black-box LLMs and exclude real SyRSs and domain experts follows from RQ1. We point out that these restrictions also ruled RAG strategies out, which would have allowed us to reduce the risk of hallucinations \cite{RAG}. Supporting the generation of SSyRSs for various domains enables broader use and more sophisticated comparisons, such as quality comparisons across domains.

To meet our objective, we designed an LLM-based generation and assessment approach for SSyRSs. It is built around a generation prompt, two assessment prompts, and iterative prompt refinements, as further described below. After ten iterations, the quality assessments of SSyRSs showed promising results. We thus designed a questionnaire study and evaluated the quality of the SSyRSs using a 33\% sample of the SSyRSs, with n=83 experts, yielding a total of n=87 submitted questionnaires.

\begin{figure*}[tb]
\centering
    \includegraphics[width=\textwidth]{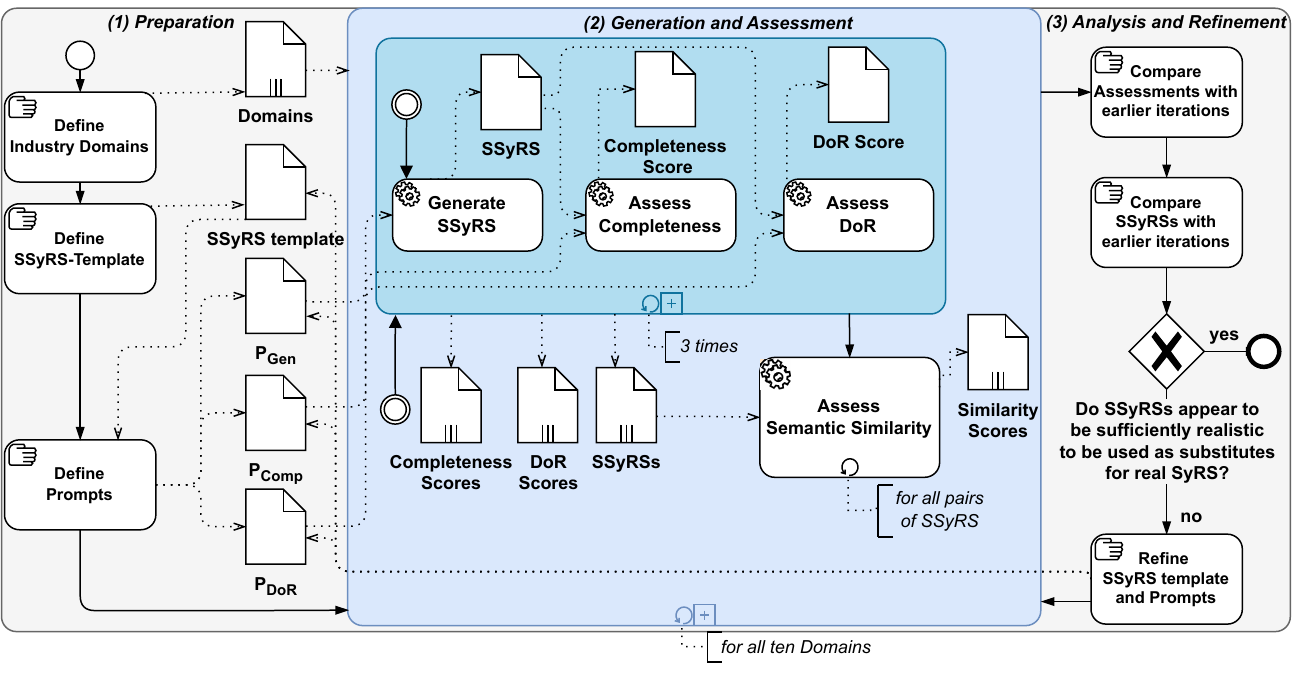}
    \caption{The SSyRS generation process (colors indicate loops, italic comments describe loop conditions).}
    \label{fig:process}
\end{figure*}

\subsection{LLM-based Generation and Assessments of SSyRSs} \label{subsec:method}

We developed an iterative generation and assessment process for SSyRSs. It is built around iterative prompt refinements based on the generation performance of earlier iterations. The process is modeled in \autoref{fig:process} using BPMN. It can be divided into three distinct phases, represented by the left, middle, and right parts of the model, briefly introduced below.

\subsubsection{Phase 1 - Preparation} We defined a set of initial data introduced in the following. Due to the iterative refinements underlying our approach, much of the data changed over time. Due to space reasons, we cannot go into detail about these changes. All versions of our data are published in Zenodo \cite{Zenodo}. Thus, we will present all data from the final iteration in the following.

\paragraph{Selected Domains:}
We generated 10 domains with a GPT-4o model using a prompt instructing it to generate domains for software products, while excluding niche domains to improve comprehensibility. Each domain was manually approved by us as a suitable candidate for our use case. The resulting domains are listed in \autoref{tbl:indDoms}.

\begin{table}[t]
\small
\centering
\caption{Industry domains the SSyRSs were generated for.}
\label{tbl:indDoms}
\renewcommand{\arraystretch}{1.2}

\begin{tabularx}{0.7\textwidth}{c X Y}
\toprule
\textbf{ID} & \textbf{Industry Domain} & \textbf{Abbreviation} \\
\midrule
1  & E-Commerce                         & \textit{e-com} \\
2  & Education                          & \textit{edu} \\
3  & Finance                            & \textit{fin} \\
4  & Government and Public Services     & \textit{gov} \\
5  & Healthcare                         & \textit{heal} \\
6  & Logistics                          & \textit{logi} \\
7  & Manufacturing                      & \textit{manu} \\
8  & Media-Entertainment                & \textit{media} \\
9  & Retail and Supply Chain            & \textit{ret} \\
10 & Telecommunications                 & \textit{tele} \\
\bottomrule
\end{tabularx}
\end{table}

\paragraph{SSyRS template:}
\begin{sloppypar}
The structure of a SSyRS should adhere to the \textit{SSyRS template} (template for short). It follows a simplified SyRS structure as defined by the ISO/IEC/IEEE 29148 standard \cite{ISO.29148}; simplified, to keep the human effort for analysis and quality assessments manageable. It comprises the overall common main categories of SyRSs. Each main category further defines a set of sub-categories to be incorporated. The four main categories and their sub-categories can be seen in the example SSyRS in \autoref{sec:results}.
\end{sloppypar}


\paragraph{Quality Properties:}
We defined three quality properties for SSyRSs to evaluate a SSyRS's suitability for its intended use from multiple angles: structural consistency, realistically modeling SyRSs, and ensuring data diversity while avoiding redundancy. They were defined as follows: 

\begin{itemize}
    \item[-] \textit{Degree of Realism (DoR):} Measures how realistic a SSyRS is in relation to real SyRSs. DoR must be ensured to enable SSyRSs to be used as substitutes for SyRSs.
    \item[-] \textit{Completeness:} Measures whether a SSyRS contains all elements defined in the template. This ensures comparability among SSyRS and consistency with the expected result.
    \item[-] \textit{Semantic Similarity:} Measures how different two SSyRSs of the same domain are. It allows for measuring the semantic overlap between multiple SSyRSs within a domain. While some overlap among SSyRSs is desirable to support comparability and data diversity, they should not be redundant. 
\end{itemize}

\paragraph{Generation Prompt:}

The core instruction of the generation prompt ($P_{Gen}$) was to generate a SSyRS according to the template for a specific domain and to vary its content from previously generated SSyRSs of this iteration to avoid creating redundant SSyRSs. The prompt referred to the data to be generated by the term ``scenario'' instead of ``SSyRS'' to reduce hallucination risks. Since GPT models are trained on internet data, we were concerned that incorrect or vague uses of the term ``SyRS'' in the internet could encourage hallucinations; a risk that would be mitigated by using a more generic instruction. We tested the performance of both terms and could not detect any clear differences between the SSyRSs, since the generated data was strictly aligned to the provided template in both cases, making the used term seemingly obsolete. Nonetheless, we still decided to use the generic term, for the above mentioned reasons, and as no disadvantages in our tests could be observed.

The core instruction was built around a set of prompt patterns and techniques as defined by Schulhoff et al. \cite{Schulhoff.2025} and White et al. \cite{White.2023}.:
\begin{sloppypar}
    \begin{itemize}
    \item[-] \textit{Zero-shot prompting}: The study design ruled the use of real SyRSs out. SSyRSs from earlier iterations were also not provided as examples, because quality assessments were limited to subjective evaluations and possibly unreliable LLM assessments. Using low-quality examples posed the risk that SSyRSs could be negatively impacted.

    \item[-] \textit{Template pattern}: Used to generate SSyRSs according to the template to ensure that SSyRSs follow a common structure. 

    \item[-] \textit{Persona pattern}: Used because its positive impact on LLM-tasks is widely known. The role was configured as an ``experienced requirements engineer and business analyst''.

    \item[-] \textit{Chain-of-thought pattern}: Used because the instructions for generating SSyRSs became overly complex; LLMs are better at solving multiple smaller problems than one complex one.
\end{itemize} 
\end{sloppypar}

\paragraph{Quality Assessments:}

We developed two assessment prompts, $P_{Comp}$ and $P_{DoR}$, to measure \textit{Completeness} and \textit{Degree of Realism} (DoR). Measuring DoR proved difficult, as the definition of when a SSyRS is considered realistic has many facets. We decided to design $P_{DoR}$ as an exploratory metric that provides a hybrid assessment by combining quantitative and qualitative elements. It instructs the LLM to calculate a \textit{DoR score} in the range $[0,1]$, where $1$ represents an entirely realistic, $0$ an entirely unrealistic SSyRS. If an unrealistic element of the SSyRS is identified, points must be deducted on the prescribed scales. The scale depends on the severity of a finding, with the severity level determined by precise instructions specifying how a finding must be classified. The sum of point deductions is then subtracted from 1, yielding the DoR score. The qualitative part is embodied in instructing the LLM to provide detailed explanations for each finding, i.e., why a finding is considered unrealistic and why it is classified by the chosen severity level. The same persona pattern as in $P_{Gen}$ is used to interpret the result from the same perspective. Moreover, chain-of-though is leveraged to reduce the complexity of the task.

$P_{Comp}$ is composed of the instruction to identify any missing content prescribed by the template, measured as a boolean. If any missing element of the template is identified, the result is false, i.e., the  SSyRS does not completely follow the template, otherwise the result is true. No prompt patterns were used for this prompt, as it performed perfectly on each assessment; prompt refinements addressed only the verbosity of responses.

To measure \textit{Semantic Similarity}, Sentence Transformers (SBERT) rather than ChatGPT are used to reduce the risk of hallucination. SBERT is capable of detecting semantic similarities with high accuracy while not incorporating the risks of hallucinations \cite{SBERT.2019}.

\subsubsection{Phase 2 -  Generation and Assessment:}
Domain-wise, in one context, we prompted ChatGPT-4o with $P_{Gen}$ to generate three SSyRSs per domain. We decided on three SSyRSs to ensure a comparable and sufficiently diverse set of SSyRSs for each domain while keeping the human effort required for result analysis manageable. After each SSyRS generation, we prompted ChatGPT-4o in the same context with $P_{Comp}$ and $P_{DoR}$ to measure Completeness and DoR of the generated SSyRS. After three SSyRSs per domain were generated and their Completeness and DoR were measured, we calculated a semantic similarity score for each pair of the three SSyRSs per domain with SBERT.

Consequently, we opted for LLM self-assessments rather than an LLM-as-a-judge approach, since the results of the self-assessments in early iterations were promising, allowing us to keep procedural overhead low by relying on a single LLM rather than multiple models. To ensure external validity due to model-specific bias, we also conducted cross-model checks on artifacts from the final iteration using Sonnet 4.5. The results of these checks, as well as potential effects of using self-assessments instead of an LLM-as-a-judge setup and possible bias introduced by shared-context usage, are discussed later.

\subsubsection{Phase 3 -  Analysis and Refinement:}
The main objective of this phase was to decide whether the SSyRSs and quality assessments show both good and realistic results compared to previous iterations. If so, the process was terminated. If not, the data was refined. There are three reasons for this subjective termination condition: 1.) LLMs are probabilistic by nature \cite{Bass.2025}. Since prompt refinements cannot guarantee improvements in the generated data, results from an iteration must be considered alongside previous results to understand overall performance. 2.) The reliability of LLM judgments is a huge concern \cite{Schroeder.2025}. We were unsure whether LLM-based assessments would yield reliable judgments of the artifacts' quality. 3.) We had no comparison data for our metrics. Defining thresholds for the scores as termination conditions was, thus, infeasible. Instead, we had to consider and interpret these results in their overall contexts.

Analysis and refinement were conducted by the first two authors. Completeness scores were manually reviewed to identify incomplete SSyRSs and assessment errors. For the analysis of the DoR and similarity scores, we computed descriptive statistics (mean, median, range, standard deviation). These were analyzed in two phases: (1) a cross-domain comparison to assess comparability with prior iterations and identify domain-specific patterns by contrasting statistics with previous results; and (2) within-domain analyses to detect inconsistencies and outliers. As a final step, all SSyRSs were rated by the authors on a subjective scale of how realistic they appeared.

\subsection{Process Execution}

The process was executed from January 13, 2025, to January 25, 2025. We opted for ChatGPT over the OpenAI API due to lower costs, as our tasks involved generating many tokens, and for the 4o models, as they were outperforming other GPT models at the time of the study. To calculate similarity scores, we used SBERT's pre-trained \texttt{all-mpnet-base-v2} model with cosine similarity due to high accuracy across various semantic textual similarity tasks \cite{SBERT.2019}. A similarity score of 1 is interpreted as ``semantically 100\% overlapping'', a score of 0 as ``semantically 100\% distinct''. 

Our termination criterion was met after ten iterations. First, we indeed observed positive effects on generation and evaluation performance, which, based on our analyses, were attributable to the prompt refinements of $P_{Gen}$ and $P_{DoR}$. However, these effects diminished in later iterations, which means that either our changes did not lead to better results or we could not attribute the results on average (exclusively) to the refinements, so that the improved results appeared to be rather random than controlled. Secondly, we obtained a large dataset of $300$ SSyRSs for further analysis. Lastly, the SSyRSs from the tenth iteration showed the best results for the DoR scores, similarity scores, and our subjective assessments. Thus, we considered these SSyRSs as suitable for expert evaluation.

\section{\uppercase{Results}} \label{sec:results}

In total, we generated 300 SSyRSs, with 3 SSyRSs per domain in each iteration. All SSyRSs are published in our dataset \cite{Zenodo}. Since the dataset of the tenth iteration passed our termination criterion and was evaluated by experts, we will present the results of that iteration in the following. We first share insights into size and quality statistics, then provide an excerpt of an example SSyRS.

\subsection{Size Statistics}
\autoref{tab:ssyrs-size-statistics} summarizes the size values of the three SSyRSs per domain, measured in number of words. Overall, the SSyRSs have a mean length of 716 words and a median of 719 words, indicating a balanced distribution across the SSyRSs per domain. The mean size per domain ranges from a minimum of 676 words in e-commerce to a maximum of 792 words in education, indicating a moderate variability in size across the domains. Education represents the domain with the overall largest corpus of 2377 words across the SSyRSs, e-commerce the domain with the smallest corpus of 2029 words. Both domains also contain the overall largest SSyRS with 811 words and the smallest one with 644 words; on average, the size shows a low overall variance. In total, the dataset comprises 21,478 words. In terms of size, the dataset, thus, represents a substantial and overall homogeneous corpus.

\begin{table}[t]
\small
\centering
\caption{SSyRS size values (measured in number of words).}
\label{tab:ssyrs-size-statistics}
\renewcommand{\arraystretch}{1.2}

\begin{tabularx}{\textwidth}{lYYYYY}
\toprule
\textbf{Domain} & \textbf{Mean} & \textbf{Median} & \textbf{Min} & \textbf{Max} & \textbf{Total} \\
\midrule
e-com  & 676 & 661 & 644 & 724 & 2029 \\
edu    & 792 & 802 & 764 & 811 & 2377 \\
fin    & 713 & 715 & 695 & 728 & 2138 \\
gov    & 688 & 676 & 663 & 724 & 2063 \\
heal   & 725 & 720 & 705 & 751 & 2176 \\
logi   & 692 & 707 & 644 & 726 & 2077 \\
manu   & 707 & 715 & 687 & 718 & 2120 \\
media  & 687 & 683 & 651 & 726 & 2060 \\
retail & 757 & 761 & 726 & 785 & 2272 \\
tele   & 722 & 732 & 693 & 741 & 2166 \\
\midrule
\textbf{Overall} & \textbf{716} & \textbf{719} & \textbf{644} & \textbf{811} & \textbf{21,478} \\
\bottomrule
\end{tabularx}
\end{table}

\subsection{Quality Statistics}
As the completeness scores were valid for all SSyRS, we discuss statistics on the semantic similarity and DoR scores in the following. 

\autoref{tbl:semantics} depicts descriptive statistics on the semantic similarity scores of the dataset. Overall, the scores have a mean of 0.66 and a median of 0.67, indicating a moderate level of semantic similarity within the dataset. However, across the domains, the mean scores range from 0.59 (government) to 0.73 (healthcare), indicating significant but moderate variation in semantic similarity. The dispersion of scores differs depending on the domain, with standard deviations ranging from 0.01 (e-commerce), indicating very consistent similarity scores in this domain, to 0.12 (manufacturing), indicating greater variance. The lowest overall similarity score is 0.50 (finances), while 0.82 (logistics) represents the highest. This indicates that some domains tend to have more variance in their
SSyRSs than others. Overall, these results indicate that the dataset represents a heterogeneous set of SSyRSs, containing similar yet not entirely redundant data, ensuring a high level of diversity across the domains.

As our analyses revealed that DoR scores from the DoR assessments were unreliable, we conducted additional validation tests to better interpret the scores and to ensure external validity by understanding possible context-related and model-related biases. The settings were:  
\begin{itemize}
    \item \textit{GPT-4o (same context)}: measured DoR scores in the same context in which the SSyRSs had been generated.
    \item \textit{GPT-4o (new context)}: measured DoR scores in a new context. It was chosen to examine the effects of context-related biases on scores.
    \item \textit{GPT-5.2 (instant)}: was chosen as being a very recent model, allowing comparisons to be made between older and new GPT models.
    \item \textit{GPT-5.2 (think)}: was chosen as being the most advanced GPT reasoning model at the time writing.
    \item \textit{Sonnet 4.5}: was chosen to understand model-related biases, and because it performs better than GPT on most coding-related tasks; this could be an advantage over GPT models, given the technical nature of SSyRSs.
\end{itemize}


The validation tests were conducted on January 22, 2026. All models were used via their respective chat interfaces. \autoref{tbl:dor} depicts the results. It shows the average DoR scores across domains and model settings. Overall, the results indicate consistently high realism assessments, with domain-related mean DoR scores ranging from 0.77 (healthcare) to 0.82 (manufacturing). Moreover, most DoR scores fall within 0.79–0.81, indicating that domain-related effects on the DoR scores are low. In contrast, comparing the model settings reveals significant effects on the DoR scores. While the average scores of \textit{GPT-4o (new context)} (0.86) and\textit{ GPT-5.2 (instant)} (0.85) are slightly lower than those of \textit{GPT-4o (same context)} (0.9), the differences are little. This indicates that although the DoR scores that were used in our dataset may have been influenced by context-related bias, the influence is negligible. However, a noticeable drop is observed in \textit{GPT-5.2 (think)} (0.73), indicating more conservative realism assessments. \textit{Sonnet 4.5} was by far the most critical model (0.64), as it recorded the lowest average values in each measurement, where the lowest average DoR score (0.6 for healthcare) is 0.27 points lower than the lowest DoR score of \textit{GPT-4o (same context)} (0.87 for e-commerce) and even 0.08 points lower than the lowest DoR score of \textit{GPT-5.2 (think)}. These results suggest that DoR assessments are highly model-specific, with the chosen model exerting a much stronger effect than the context used.

\begin{table}[t]
\small
\centering
\caption{SSyRSs semantic similarity scores}
\label{tbl:semantics}
\renewcommand{\arraystretch}{1.15}
\begin{tabularx}{\textwidth}{YYYYYYY}
\toprule
\textbf{Domain} &
\textbf{Mean} &
\textbf{Median} &
\textbf{Std.\ Dev.} &
\textbf{Min} &
\textbf{Max} \\
\midrule
e-com  & 0.67 & 0.66 & 0.01 & 0.65 & 0.68 \\
edu    & 0.62 & 0.60 & 0.05 & 0.58 & 0.67 \\
fin    & 0.60 & 0.59 & 0.10 & 0.50 & 0.71 \\
gov    & 0.59 & 0.58 & 0.06 & 0.53 & 0.65 \\
heal   & 0.73 & 0.75 & 0.05 & 0.67 & 0.76 \\
logi   & 0.71 & 0.67 & 0.09 & 0.65 & 0.82 \\
manu   & 0.67 & 0.71 & 0.12 & 0.54 & 0.78 \\
media  & 0.62 & 0.60 & 0.04 & 0.58 & 0.67 \\
retail & 0.70 & 0.71 & 0.09 & 0.62 & 0.79 \\
tele   & 0.71 & 0.67 & 0.08 & 0.66 & 0.80 \\
\midrule
\textbf{Overall} & \textbf{0.66} & \textbf{0.67} & \textbf{0.08} & \textbf{0.50} & \textbf{0.82} \\
\bottomrule
\end{tabularx}
\end{table}

\begin{table}[t]
\small
\centering
\caption{Mean values of SSyRS DoR scores per domain and model setting}
\label{tbl:dor}
\renewcommand{\arraystretch}{1.2}

\begin{tabularx}{\textwidth}{YYYYYYY}
\toprule
\thead{Domain} &
\thead{GPT-4o\\same context} &
\thead{GPT-4o\\new context} &
\thead{GPT-5.2\\instant} &
\thead{GPT-5.2\\thinking} &
\thead{Sonnet 4.5} &
\thead{Mean DoR\\per Domain} \\
\midrule
e-com   & 0.87 & 0.85 & 0.85 & 0.74 & 0.62 & \textbf{0.79} \\
edu     & 0.90 & 0.86 & 0.85 & 0.70 & 0.61 & \textbf{0.78} \\
fin     & 0.89 & 0.86 & 0.88 & 0.74 & 0.63 & \textbf{0.80} \\
gov     & 0.91 & 0.86 & 0.86 & 0.79 & 0.64 & \textbf{0.81} \\
heal    & 0.91 & 0.83 & 0.82 & 0.68 & 0.60 & \textbf{0.77} \\
logi    & 0.89 & 0.87 & 0.89 & 0.76 & 0.62 & \textbf{0.81} \\
manu    & 0.91 & 0.87 & 0.82 & 0.73 & 0.75 & \textbf{0.82} \\
media   & 0.90 & 0.87 & 0.88 & 0.71 & 0.63 & \textbf{0.80} \\
retail  & 0.90 & 0.88 & 0.86 & 0.74 & 0.63 & \textbf{0.80} \\
tele    & 0.90 & 0.85 & 0.82 & 0.76 & 0.64 & \textbf{0.79} \\
\midrule
\textbf{Mean DoR per Model} &
\textbf{0.90} &
\textbf{0.86} &
\textbf{0.85} &
\textbf{0.73} &
\textbf{0.64} & \\
\bottomrule
\end{tabularx}
\end{table}

Moreover, some \textit{Sonnet 4.5} assessments were noticeably low, such as the DoR score of the second healthcare SSyRS, which was rated at only 0.48 by the model. To better understand the reliability of this assessment, we evaluated the SSyRS nine additional times using that model. The results are depicted in the Tables \ref{tab:healthcare-1} and \ref{tab:healthcare-2}. The DoR scores range from a minimum of 0.48 to a maximum of 0.73, i.e., 0.25.  Given the $[0,1]$ scale of the scores, the values, thus, span 25\% of their scale, indicating very unreliable assessments. On average, \textit{Sonnet 4.5} rated the DoR score at 0.59, i.e., 0.11 points better than in the first assessment, and its ratings have a standard deviation of 0.08. Considering the first quartile, after which 25\% of the DoR scores are 0.52 or less, the 0.48 rating is indeed an outlier. Overall, these results indicate that quantitative DoR assessments are highly unreliable.

\begin{table*}[t]
\small
\centering
\renewcommand{\arraystretch}{1.2}

\begin{minipage}[t]{0.48\textwidth}
\centering
\captionof{table}{DoR scores per run measured by Sonnet 4.5 for 2nd healthcare SSyRS (10 runs).}
\label{tab:healthcare-1}
\begin{tabularx}{\textwidth}{cY}
\toprule
\textbf{Run} & \textbf{DoR Score} \\
\midrule
1  & 0.48 \\
2  & 0.61 \\
3  & 0.56 \\
4  & 0.53 \\
5  & 0.60 \\
6  & 0.70 \\
7  & 0.57 \\
8  & 0.73 \\
9  & 0.58 \\
10 & 0.49 \\
\bottomrule
\end{tabularx}
\end{minipage}
\hfill
\begin{minipage}[t]{0.48\textwidth}
\centering
\captionof{table}{Descriptive statistics of DoR scores (Sonnet 4.5, 2nd healthcare SSyRS, 10 runs).}
\label{tab:healthcare-2}
\begin{tabularx}{\textwidth}{lY}
\toprule
\textbf{Statistic} & \textbf{Value} \\
\midrule
Mean       & 0.59   \\
Std.\ Dev. & 0.08   \\
Variance   & 0.0066 \\
Min        & 0.48   \\
Max        & 0.73   \\
Range      & 0.25   \\
Q1         & 0.52   \\
Q3         & 0.63   \\
\\
\\
\bottomrule
\end{tabularx}
\end{minipage}

\end{table*}

\subsection{Exemplary SSyRS} 
\autoref{fig:ssyrs-example} depicts an excerpt of the SSyRS ``Dynamic Freight Optimization Platform (DFOP)'' of the logistics domain as example. We chose this SSyRS due to quality and space reasons. With a DoR score of 0.85 across the models, it has the highest DoR score of all SSyRSs. Moreover, it is the smallest SSyRS in the dataset, allowing us to share the most complete excerpt relative to its original size. The excerpt contains 461 of the 644 words, i.e., is of $\sim$72\% of the size of the original SSyRS and presents all main categories and sub-categories as prescribed by the SSyRS template, as well as at least one example data per sub-category. All data is completely generated by ChatGPT.

\begin{figure*}[p]
\centering
\begin{tcolorbox}[
  width=\linewidth,
  colback=gray!3,
  colframe=gray!60,
  boxrule=0.6pt,
  arc=2pt,
  left=6pt,right=6pt,top=6pt,bottom=6pt
]
\scriptsize

\textbf{1. System Overview}\par

\textbf{System Purpose.} The system, called "Dynamic Freight Optimization Platform (DFOP)," aims to optimize freight management for a multinational logistics company by providing real-time freight tracking, dynamic route optimization, load consolidation, and predictive analytics for demand forecasting. The goal is to reduce operational costs, improve delivery times, and enhance overall supply chain efficiency.
\par
\textbf{Domain/Context:
} The logistics domain, specifically targeting freight management in intermodal transport, spanning road, rail, and maritime networks.
\par

\textbf{Stakeholders:}
\begin{itemize}[leftmargin=*,nosep]
  \item Internal: Operations managers, dispatchers, warehouse personnel
  \item External: Freight carriers, logistics partners, end customers
  \item Regulatory Authorities: To ensure compliance with transport and trade regulations.
\end{itemize}

\textbf{User Base Characteristics:}
\begin{itemize}[leftmargin=*,nosep]
  \item Size: $\sim$20{,}000 active users, including logistics coordinators, drivers, warehouse staff, and customers across 30 countries from North America, Europe, and Asia, with global access.
  \item Roles: Dispatchers, fleet managers, drivers, warehouse staff, customers (tracking orders), and analysts.

\end{itemize}

\textbf{Operational Environment:} 
\begin{itemize}[leftmargin=*,nosep]
    \item Hosting: Cloud-based infrastructure deployed on a multi-region architecture
    \item Access: Mobile app for drivers, web-based platform for administrators, APIs for partners
\end{itemize}

\textbf{Usage Scenarios:}
\begin{itemize}[leftmargin=*,nosep]
  \item Dispatchers schedule and monitor shipments via the system.
  \item Drivers receive optimized routes and report delivery statuses through the mobile app.
  \item Customers track shipments and view estimated arrival times in real time.
\end{itemize}

\vspace{4pt}

\textbf{2. Functional Requirements}\par
\textbf{Core Features:}
\begin{itemize}[leftmargin=*,nosep]
    \item Real-Time Freight Tracking: GPS-based shipment tracking with status updates.
    \item Dynamic Route Optimization: Automated routing based on traffic, weather, and priorities.
    \item Load Consolidation: Suggestions for combining shipments to optimize truck capacity.
    \item Predictive Analytics: AI-driven demand forecasting for peak periods and supply planning.
\end{itemize}

\textbf{Authentication Conditions \& Frequency:}\par

\begin{itemize}[leftmargin=*,nosep]
    \item  Session Expiration: For web and mobile platforms, expiration after 8 hours of inactivity.
    \item Sensitive Actions: Authentication required for creating shipment schedules, altering delivery routes, or accessing predictive analytics reports.
\end{itemize}

\textbf{Sensitivity of Actions \& Permission Levels:}\par

\begin{itemize}[leftmargin=*,nosep]
    \item Dispatchers can modify delivery routes or assign shipments.
    \item Drivers can only view assigned routes and report statuses.
    \item Administrators can access all analytics and system configurations.
\end{itemize}

\vspace{4pt}

\textbf{3. Non-Functional Requirements}

\textbf{Performance:}
\begin{itemize}[leftmargin=*,nosep]
    \item Handle up to 10,000 concurrent users with response times under 2 seconds for 95\% of traffic.
\end{itemize}

\textbf{Scalability:}
\begin{itemize}[leftmargin=*,nosep]
    \item Scalable to support a 50\% increase in shipment volume during peak seasons.
\end{itemize}

\textbf{Reliability:}
\begin{itemize}[leftmargin=*,nosep]
    \item System availability of 99.9\%, with automatic failover mechanisms across regions.
\end{itemize}

\textbf{Security:}
\begin{itemize}[leftmargin=*,nosep]
    \item Encrypted data in transit and at rest using industry-standard protocols.
\end{itemize}

\textbf{Usability:}
\begin{itemize}[leftmargin=*,nosep]
    \item Mobile app with offline functionality for drivers in areas with limited connectivity.
\end{itemize}

\textbf{Audit \& Monitoring:}
\begin{itemize}[leftmargin=*,nosep]
    \item Daily audit reports for compliance and system integrity checks.
\end{itemize}

\vspace{4pt}

\textbf{4. Constraints}

\textbf{Technical Constraints:}
\begin{itemize}[leftmargin=*,nosep]
    \item Integration with third-party GPS providers and legacy fleet management systems.
\end{itemize}

\textbf{Compliance Requirements:}
\begin{itemize}[leftmargin=*,nosep]
    \item GDPR compliance for personal data of customers and employees.
\end{itemize}

\textbf{Resource Constraints:}
\begin{itemize}[leftmargin=*,nosep]
    \item Budget: \$12 million for development and deployment over 18 months.
\end{itemize}

\textbf{Integration Needs:}
\begin{itemize}[leftmargin=*,nosep]
    \item Must integrate seamlessly with existing ERP systems and external partner APIs.
\end{itemize}

\end{tcolorbox}
\caption{Excerpt of the logistics SSyRS ``Dynamic Freight Optimization Platform (DFOP)''. The whole SSyRS can be found in our GitHub repository.}
\label{fig:ssyrs-example}
\end{figure*}


\section{\uppercase{Evaluation}} \label{sec:Evaluation}
We conducted a questionnaire study to evaluate the SSyRSs from the final iteration. To this end, we defined the central question to be answered by the evaluation: ``Are SSyRSs realistically rated by ChatGPT perceived as realistic by software experts, too?'' 

\subsection{Evaluation Setup}

\subsubsection{Questionnaire Design}
We implemented the questionnaire with the SoSci Survey platform \cite{soscisurvey}. From the 30 SSyRSs, we randomly selected one SSyRS per domain and designed one questionnaire for each of these SSyRSs. At the beginning of the questionnaire, participants specified their industry and domain background. Depending on the answer, participants were assigned only the questionnaire of the corresponding domain. If participants selected multiple domains, the questionnaire was randomly assigned to one of them. After completing the questionnaire, they could choose another questionnaire from their selected domains or end the survey. If there was no match with the industry background or if the participant did not consider themselves to be a specialist or domain expert, the questionnaire was ended. The questionnaire followed the four main SSyRS categories. Participants were asked to evaluate these categories individually, using two answer options: “realistic” and “unrealistic”. Participants were explicitly informed that these options were meant to reflect tendencies rather than absolute correctness. Each assessment included an optional free-text field for comments on the rating. In the end, participants gave an overall realism assessment on a five-point Likert scale, ranging from \textit{very artificial} to \textit{very realistic}.

\subsubsection{Participant Selection}
Participants were recruited using purposive sampling and snowball sampling. The questionnaire was advertised via personal contacts, LinkedIn, and the mailing list of the ``Software Engineering'' special interest group of the German Informatics Society (GI). We explicitly addressed individuals with professional experience in software engineering. Participation was voluntary and required informed consent, which outlined key aspects of the study, including data protection measures, the purpose of data collection, the specific data gathered, and how the data would be used. Participants were informed that their responses would be anonymized and used solely for research purposes. A contact e-mail was provided for any questions or concerns regarding data usage. No incentives were offered. To ensure valid responses despite the open distribution of the questionnaire link, we included filter questions at the start to exclude participants without professional software experience or without confidence in assessing a SSyRS in at least one domain. We also asked about their years of experience and whether they identified as technical experts, domain experts, or both. Table \ref{tbl:participant-demographics} summarizes the demographics of the 83 participants. The majority of experts had over 7 years of experience, and approximately 58\% identified as having domain expertise.

\subsubsection{Implementation}
A pretest was conducted with five participants, each of whom freely selected one of the ten available SSyRSs. The participants were personal contacts from our network. As a result, the pretest covered four different domains, with two participants selecting the same domain. We measured average completion time and collected qualitative feedback to improve the questionnaire's clarity and usability. Based on the feedback, we added brief guidelines clarifying that binary judgments should reflect tendencies rather than absolute correctness. Participants were also instructed to evaluate only the presented content. The final study was conducted between March 26 and June 15, 2025. 

\subsubsection{Data Analysis}
We exported all complete and incomplete datasets from SoSci Survey without any modification. Numeric codes used by SoSci Survey (e.g., for Likert scale responses) were mapped back to their corresponding textual labels to facilitate interpretation. The data was parsed and analyzed using the Python library \texttt{pandas} and \texttt{matplotlib.pyplot} for visualization.

\subsection{Evaluation Results}

\begin{table}[t]
\small
\centering
\caption{Participant demographics (N=83)}
\label{tbl:participant-demographics}
\begin{tabular}{lcc}
\toprule
\textbf{Category} & \textbf{Group} & \textbf{Percentage} \\
\midrule
\multirow{5}{*}{Years of Experience} 
& $<$1 year & 8.4\% \\
& 1–3 years & 16.9\% \\
& 4–6 years & 13.3\% \\
& 7–10 years & 50.6\% \\
& $>$10 years & 10.8\% \\
\midrule
\multirow{3}{*}{Expert Type} 
& Technical & 42.2\% \\
& Domain & 19.3\% \\
& Both & 38.6\% \\
\bottomrule
\end{tabular}
\end{table}

\autoref{tbl:participant-demographics} shows the demographics of the study. 83 participants participated in our study, of which almost 75\% had 4 years of experience or more. Technical experts accounted for just under half of the participants, while domain experts accounted for around 20\%. $\sim39\%$ percent of participants stated that they were both technical and subject matter experts.

\begin{figure}[b]
    \centering
    \includegraphics[width=0.6\linewidth]{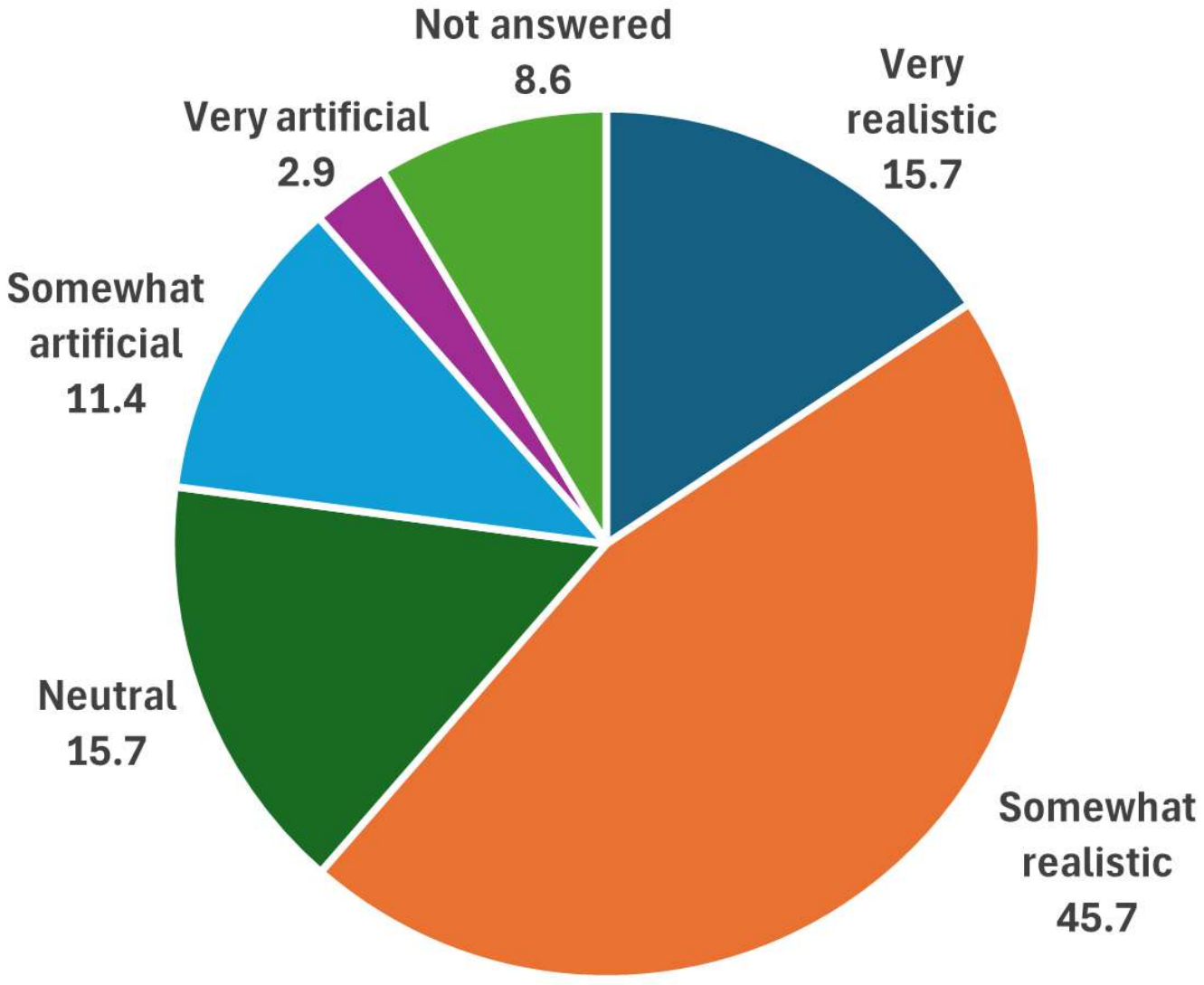}
    \caption{Distribution of the overall rating of the degree of realism of the SSyRSs.}
    \label{fig:overall-realism}
\end{figure}

Of the 83 participants, 87 questionnaires were collected; four participants submitted two questionnaires. 70 of the 87 were completely answered. A total of 227 comments were collected. All datasets were included in the analysis, as each SSyRS section was evaluated independently, and partial responses did not introduce analytical bias. Figure \ref{fig:overall-realism} presents the overall rating. Combined, 61.4\% of experts rated their SSyRS as either somewhat or even very realistic. In comparison, only 14.3\% gave a rating of either somewhat or very artificial.

To complement the quantitative analysis and put the estimations into context, we examined the optional justification comments provided for sections marked as ``unrealistic''. In these responses, four recurring groups of issues were identified: 1) Oversimplification and generic phrasing indicated by comments like ``Look like textbook examples'' and ``buzzword bingo''; 2) Lack of detail and ambiguity described as ``That is very vague. [...] more information [...] should be defined.''; 3) Logical incoherence between requirements like a misapplication of EU standards to non-EU settings; 4) Overly ambitious requirements, such as ``15.000 users for an initial round seems too ambitious''.

Overall, the data analysis provides a nuanced answer to the question posed in this evaluation study. Overall, the SSyRS appear to be realistic to experts, as over 60\% of experts gave an overall rating of ``somewhat realistic'' or even ``very realistic''. However, a deeper analysis of the expert comments shows that this assessment may not accurately reflect the DoR of the sample. Experts who conducted an in-depth analysis, as evidenced by their comments and the time spent on the evaluation, identified some major shortcomings, particularly in the areas of \textit{Resource Constraints}, \textit{User Base Characteristics}, and non-functional requirements such as \textit{Scalability}, \textit{Reliability}, and \textit{Performance}, which were only apparent upon closer inspection.


\section{Discussion}
Our results provide differentiated insights into using ChatGPT for generating realistic SSyRS. We discuss them by formulating the most important insights we have gained.

\begin{tcolorbox}[colframe=black, boxrule=0.5pt, arc=0mm, left=2mm, right=2mm, top=1mm, bottom=1mm]
\normalsize
\textbf{Insight 1:} Prompt-based structural validation of complex SSyRSs generated using the template prompting pattern is very effective. Semantic similarity is an effective metric for ensuring diversity in generated SSyRSs.
\end{tcolorbox}

Central to our approach was the use of a template to generate SSyRSs and a corresponding prompt ($P_{Comp}$) to validate whether each SSyRS fully implements the template. Our results show that this approach is highly reliable, as the assessments by $P_{Comp}$ were consistently correct throughout all iterations. Moreover, the semantic similarity metric proved effective for ensuring diversity in SSyRSs. Consequently, we can recommend both approaches for evaluating the quality of complex LLM-generated documents in our setting.

\begin{tcolorbox}[colframe=black, boxrule=0.5pt, arc=0mm, left=2mm, right=2mm, top=1mm, bottom=1mm]
\normalsize
\textbf{Insight 2:} LLM-based quantitative plausibility measurements, such as the degree of reality (DoR) of SSyRSs, are unreliable. Qualitative assessments are an effective method for identifying plausibility weaknesses.
\end{tcolorbox}

We designed a hybrid assessment prompt $P_{DoR}$ to measure both quantitatively and qualitatively how realistic a SSyRS is in relation to real SyRSs. Our results indicate a high degree of unreliability in quantitative assessments. While we did not observe a significant impact of context-based biases on reliability, model-based biases pose a serious problem. Evaluations based on individual measurements do not provide a reliable basis for plausibility checks, as the measured values show a very high degree of variation. Reliability can be improved by calculating average values based on multiple measurements. However, even average values do not provide a reliable basis, as the evaluations are overly dependent on the evaluating LLM. According to our results, evaluations by LLMs with strong reasoning performance, such as the GPT5.2 Thinking Model or Sonnet 4.5, are more reliable because the evaluations are significantly more critical on average, which is consistent with our expert evaluation results. However, further studies are needed to investigate this in detail. Until then, we must advise against relying on quantitative plausibility assessments of complex synthetic RE artifacts, making manual evaluation by experts indispensable. In contrast, the qualitative assessments were very valuable in identifying contradictory or unrealistic elements. Across all LLMs, there was a high degree of overlap in the findings identified, and the explanations provided helped us refine the generation prompt; only the number of points deducted by the LLMs for the findings showed a high degree of variance, which also explains the unreliability of quantitative assessments.

\begin{tcolorbox}[colframe=black, boxrule=0.5pt, arc=0mm, left=2mm, right=2mm, top=1mm, bottom=1mm]
\normalsize
\textbf{Insight 3:} Expert analyses of SyRSs are prone to error and require a high degree of analytical diligence due to the technical complexity and LLM-typical presentation of the data. 
\end{tcolorbox}

Given that SSyRSs are not generated as realistically as they seem to us (and ChatGPT), why do most experts still consider the sample realistic? We see three key factors that contributed to this observation:

\begin{itemize}[nosep]
    \item[i)] \textbf{LLM-based overconfidence}. The confident and structured presentation of LLM-generated text likely creates an illusion of realism, particularly in NL artifacts. Since LLMs are trained primarily on public internet data, often lacking the nuanced detail found in expert-level documentation, they tend to generalize or hallucinate seemingly plausible content that may not withstand domain scrutiny.
    \item[ii)] \textbf{Artifact complexity}. Given the average SSyRS size of 716 words, they are rather complex NL artifacts. Later elements in a SSyRS may contradict earlier parts of it; multiple flaws were caused by such contradicting parts across the SSyRSs. It is not a trivial task to analyze complex NL artifacts in such detail as to allow the identification of unrealistic or contradictory parts in a SSyRS. The LLM-typical presentation style and overconfidence exacerbate this issue further.
    \item[iii)] \textbf{Attitudes toward AI}. The participating experts may have been influenced by their attitudes toward AI. Research on AI appreciation and aversion shows that trust in AI is shaped by perceived capability, personalization, and social influence \cite{Klingbeil.2024}. Knowing upfront that the SSyRSs were AI-generated could have triggered both positive bias (novelty or trust in automation) and negative skepticism (distrust or critical framing), depending on individual predispositions.
\end{itemize}

\vspace{10pt}

\noindent Does this mean that the SSyRSs we have generated are unusable? No, they are not. The SSyRSs resemble real SyRSs to a high degree in terms of structure and content, which most experts also agree on. However, a distinction must be made as to whether the dataset is free of errors, and this is not the case. Nevertheless, the dataset forms a valuable basis for manual quality analysis and refinement by experts, as well as for the extraction of context-embedded system requirements for future work. We conclude this paper with an answer to our central research question. 

\vspace{5pt}

\begin{tcolorbox}[colframe=black, boxrule=0.5pt, arc=0mm, left=2mm, right=2mm, top=1mm, bottom=1mm]
\normalsize
ChatGPT is a powerful tool for generating SSyRS candidates and providing a rough initial assessment of their degree of realism without access to real SyRSs. However, a thorough, careful review by experts is essential to ensure that SSyRS candidates are actually realistic SSyRSs. A fully automated evaluation of SSyRS cannot be conducted without expert input, as LLM-based evaluations are unreliable.
\end{tcolorbox}

\subsection{Limitations}

The template we used to generate SSyRSs was synthetic and simplified. It might not capture the full complexity of SyRSs and can influence how realistic the SSyRSs are perceived. To minimize this threat, we carefully designed the template to ensure that all included parts complied with an official standard \cite{ISO.29148}. Only technically complex details that are difficult to add plausibly without a specific system at hand, such as existing interface definitions, were omitted.

We observed improvements in the semantic similarities and in our subjective assessments throughout the iterations. While the prompt refinements may have led to these improvements, the probabilistic nature of LLMs cannot be ruled out as a possible cause. Moreover, all prompts were improved based on objective criteria and subjective evaluations, complicating causal interpretation. We deliberately decided against a completely objective approach due to the sheer complexity of the design space. Not only are there numerous prompt patterns to consider, but the effect of each pattern can vary significantly depending on its specific wording and contextual integration. To reduce this threat, we avoided general causal relationships between prompt changes and their effects on results and took an exploratory approach throughout the study.

Some limitations are induced by our expert study. We performed purposive sampling with additional snowball sampling. This approach does not allow for full verification of participant expertise. Moreover, participation relied on self-reporting, which cannot be independently verified. This poses the risk that some participants may lack the expertise required for an in-depth evaluation, introducing some uncertainty regarding the consistency and reliability of the expert judgments. We tried to minimize these risks by implementing several screening questions, including filters for professional software experience and self-identification as a domain or technical expert. Moreover, participation was voluntary, and no incentives were offered. Next, the expert study involved a single SSyRS per domain. This small sample size limits statistical power and reduces the generalizability of the findings. Therefore, we did not compare the results across the individual domains and treated them as a single sample of the 30 final SSyRSs. Since these are semantically different, it is nevertheless likely that the domain choice influenced the expert assessment. This design decision allowed us to examine a larger overall sample by identifying multiple experts across many domains. In return, we had to accept the disadvantages of limited statistical power and reduced generalizability. However, we consider this a lesser evil, since the alternative would have been a smaller overall sample in one domain, which would have prevented the results from generalizing. Lastly, the case study's design poses another limitation: it was restricted to 10 industry domains we defined upfront. Consequently, the results of our study cannot be transferred to industry domains that were not included in the case study.

\section{Related Work}
Cosler et al. \cite{Cosler.2023} developed the nl2spec framework, which can translate unstructured NL requirements into formal specifications in temporal logics. Zhang et al. \cite{Zhang.2024} propose an AI-aided Software Development approach. Among other things, it takes user requirements as input and generates use cases. Iqbal et al. \cite{Iqbal.2019} developed a framework for automatically generating software requirements for new mobile applications by mining app store content and competitor feedback. Yeow et al. \cite{Yeow.2024} developed an LLM-based approach using GPT-3.5 to generate survey and interview questions for requirements elicitation across various industry domains. They assess the linguistic quality and contextual relevance of the LLM-generated questions using metrics such as Flesch Reading Ease and expert evaluation. 
Moreover, Peer et al. \cite{Peer.2024} developed the ``Natural Language Processing for Requirement Forecasting'' (NLP4ReF) framework, which utilizes a natural language toolkit and ChatGPT to classify requirements by functional and non-functional requirements as well as their respective system classes, and to generate overlooked requirements from an existing requirements set. 

These approaches share our goal, but differ from our approach in terms of their inputs, outputs, or the methods used to generate the data. Iqbal et al. \cite{Iqbal.2019} address the same problem as our approach: scenarios where real-world requirement specifications or expert input are unavailable, and to generate RE artifacts without relying on existing user data. However, the authors do not use an LLM but an ML-pipeline based on NLP and traditional ML techniques to generate data. We used a black-box LLM instead. Moreover, their approach generates only software requirements, whereas our approach generates synthetic SyRSs that add context to the requirements. Cosler et al. \cite{Cosler.2023} use existing NL requirements as input, while our approach is built on replaceable prompts, having no dependencies on available data. Moreover, the generated data of nl2spec are formal specifications in temporal logics, while we generated full specifications in NL. Zhan et al. \cite{Zhang.2024} generate use cases, while we generated SSyRSs. Moreover, AISD requires user requirements as input to generate use cases. Our approach does not depend on any historical data. Yeow et al. \cite{Yeow.2024} focus on facilitating the elicitation phase for a predefined system by automatically generating survey and interview questions for requirements elicitation. In contrast, we generated requirements, and we did this for a novel, synthetic system.

Lastly, our work is related to the PROMISE dataset, which is considered a gold standard in this field \cite{PROMISE}. The dataset covers a wide range of functional and non-functional requirements and can therefore be compared with those of our SSyRSs. However, each SSyRS defines additional and important contexts. This additional context significantly affects the possible uses of these requirements, as they can be interpreted and applied differently depending on the context.

\section{Conclusion}
This case study examines whether ChatGPT can generate realistic synthetic system requirement specifications (SSyRSs). It introduces three quality properties to be ensured when generating SSyRSs, along with metrics to calculate them without domain expert involvement. The underlying generation approach follows iterative prompt refinements based on diagnostic data from earlier iterations. We implemented and executed the approach in a case study, and generated 300 SSyRSs for ten different industry domains in total. We made all SSyRSs, prompts, and assessment results publicly available \cite{Zenodo}. After ten iterations, our quality as well as our own assessments rated the resulting SSyRSs to be mostly realistic. To evaluate their quality, we evaluated a 33\% sample of the resulting SSyRSs in an expert questionnaire study. The study revealed that over 60\% of the participants perceived the sample as realistic to very realistic, which is in line with the LLM-based assessments. However, in-depth analyses of the SSyRSs raised serious concerns about these assessments. Our results show that quantitative plausibility assessments of SSyRSs are too unreliable, due to model-related biases. Moreover, they show a dangerous interplay between human perception, LLM overconfidence, and the plausible formulations LLMs tend to generate. 

For future work, the functional and non-functional requirements can be extracted from our SSyRSs and investigated to serve as an alternative dataset to PROMISE. Moreover, we plan extensive follow-up expert evaluations to increase statistical power, for instance, by comparing expert perceptions of the SSyRSs from the first and tenth iterations. Lastly, we plan to refine the dataset with experts from their domains.

\FloatBarrier

\section*{Acknowledgements}
We thank all participants who contributed to our questionnaire study.

\bibliographystyle{unsrt}  
\bibliography{Bibliography}


\end{document}